# Quantum Theory of Amplification in RS

A.S. Badikyan, K.S. Badikyan [*], D.K. Hovhannisyan

National University of Architecture and Construction, Yerevan, Armenia

[*] badikyan.kar@gmail.com

## Abstract

The equations for the amplitudes of transition probabilities are found. The resonance frequency of the system is calculated. It is shown the dependence of the resonance frequency on the initial transverse coordinate of an electron. This dependence is analized.

1. ## Introduction

Relativistic strophotron is a system in which fast electrons move along a potential "trough" produced by quadrupole magnetic or electric lenses.

The best-known free-electron lasers among the experimentally implemented systems [1-15] are based on the use of undulators in which a static transverse magnetic field varies periodically along the direction of electron motion (*OZ* axis). A different free-electron laser is discussed in Ref. [3]: in this case a beam of relativistic electrons is traveling in a static magnetic or electric field, the potential of which is homogeneous along the direction of motion of electrons (OZ axis) and rises parabolically along one of the transverse directions (*OX* axis). The name strophotron is sometimes used for these systems [3].

The problem of amplification of an external wave is solved in Ref. [3] without full allowance for the longitudinal and transverse electron motion. We shall show that the changes in the longitudinal energy associated with the emission (or absorption) of external wave photons have a considerable influence on the frequency of the oscillations of the electrons in the field and on the nature of the wave functions describing the transverse electron motion.

A geometry of the fields similar to that considered in the present study applies also to the case of emission of radiation as a result of channeling of electrons and positrons in crystals [4]. In contrast to the conventional channeling, the interaction of electrons with an external macroscopic field (which can be described as macroscopic channeling) is much simpler because there are no such problems as the absorption of electromagnetic waves in a crystal, possible limits on the channeling length, etc.

The feasibility of practical realization of a potential of this type is considered in Ref. [6,7], when a system of specially oriented magnetic quadrupole lenses located at equal distances from one another on the beam axis is proposed [6].

We shall use a method based on the direct solution of the initial secular quantum-mechanical problem, which in this respect differs from the S-matrix approach. We shall consider only the linear approximation although the general equations obtained are valid also in the case of many-photon processes and saturation.

## 2. Equations for the amplitudes of transition probabilities

Following Ref. [3], we shall consider planar geometry and assume that a static electric or magnetic field is independent of one of the transverse coordinates, for example, $y$. The field gradient is directed along the $OX$ axis.

We shall consider an ultrarelativistic electron of energy $\varepsilon = m\gamma \gg m$ ($\hbar = c = 1$) moving in the $XOZ$ plane at a small angle $\alpha$ to the $OZ$ axis and experiencing at an initial moment t=0 a static electric or magnetic field with the potential or

$$\Phi(x) = \Phi_0(x^2/d^2) \quad \mathbf{A}(x) = \mathbf{A}_0(x^2/d^2), \tag{1}$$

where $e\Phi_0 = e|\mathbf{A}_0|$ and 2d are, respectively, the maximum height of a transverse potential barrier and the size of the region of its localization (aperture) in the direction of the $OX$ axis; the $\mathbf{A}_0$ vector is parallel to the $OZ$ axis, i.e., the magnetic field is directed along the $OY$ axis.

In the absence of an electromagnetic wave an electron experiences translational motion along the $OZ$ axis, characterized by a momentum $\mathbf{p}_\parallel$ and an energy $\varepsilon_\parallel = \sqrt{p_\parallel^2 + m^2}$, oscillating in the transverse direction (along the $OX$ axis). The oscillation wave functions $\varphi_l(x, \varepsilon_\parallel)$, satisfy the usual Schrodinger equation obtained from the initial Klein-Gordon equation ignoring the square of the potential $\Phi(x)$ (we shall now obtain estimates justifying these and other approximations):

$$d^2\varphi_l(x,\varepsilon_\parallel)/dx^2 = \left[2\varepsilon_\parallel e\Phi(x) - 2\varepsilon_\parallel \Omega(\varepsilon_\parallel)(l+1/2)\right]\varphi_l(x,\varepsilon_\parallel), \tag{2}$$

where $\Omega(\varepsilon_\parallel) = \left(2e\Phi_0/\varepsilon_\parallel\right)^{1/2}/d$ is the oscillation frequency in a parabolic well; l= 0, 1,2,... are the oscillation quantum numbers; $\varphi_l(x,\varepsilon_\parallel) = \left[\Omega(\varepsilon_\parallel)\varepsilon_\parallel\right]^{1/4} \chi_l(\varsigma)$, $\chi_l(\varsigma)$ are the usual oscillation functions dependent on the dimensionless variable $\varsigma = (\Omega\varepsilon_\parallel)^{1/2}$ and expressed in terms of Hermite polynomials [15].

We shall assume that an electromagnetic wave of frequency $\omega$ linearly polarized in the *XOZ* plane and described by the vector potential

$$A_{em} = (E_0/\omega)\cos[\omega(t-z)], \qquad (3)$$

where $E_0$ is the amplitude of the electric field in the wave, travels along the OZ axis parallel to the direction of longitudinal motion of an electron beam.

We shall represent the wave function $\Psi$ of an electron by an expansion

$$\Psi = \sum_n C_n(t,x)\exp\{i[(p_\square + n\omega)z - (\varepsilon_\square + n\omega)t]\}, \qquad (4)$$

where $n = 0, +1, +2,...$ and the exponential factor allows explicitly for the change in the longitudinal energy and the corresponding momentum as a result of the absorption (n<0) or stimulated emission (n>0) of |n| quanta of the wave (3).

The coefficients $C_n(t,x)$ can be expanded in terms of any complete system of functions dependent on x. The most convenient basis is a system of functions $\varphi_l^{(n)}(x) \equiv \varphi_l(x,\varepsilon_\square + n\omega)$, which satisfy Eq. (2) with a shifted energy $\varepsilon_\square + n\omega$ f and correspond to a shifted oscillation frequency $\Omega^{(n)} = \Omega(\varepsilon_\square + n\omega)$:

$$C_n(t,x) = \sum_l (-1)^l a_{nl}(t)\varphi_l^{(n)}(x). \qquad (5)$$

The coefficients $a_{nl}(t)$ represent the amplitudes of the probability of finding an electron at the l th oscillation level on absorption (emission) of \n\ quanta of the field (3). The equations for $a_{nl}(t)$ follow from the Klein-Gordon equation where we can ignore terms proportional to the quantities $\Phi^2$, $A_{em}^2$, $d^2 a_{nl}/dt^2$, and $\Phi(da_{nl}/dt)$ (for estimates see below):

$$i\frac{da_{nl}}{dt} + \left[\frac{n\omega}{2\gamma^2} - (l+1/2)\Omega - \frac{(n\omega)^2}{2\varepsilon_\square\gamma^2} + \frac{n\omega(l+1/2)\Omega}{2\varepsilon_\square} - \frac{3}{8}\frac{(n\omega)^2(l+1/2\Omega)}{2\varepsilon_\square^2}\right] a_{nl}$$

$$= i\frac{eE_0}{2\varepsilon_\square\omega}\sum_{l'}(-i)^{l'-l}\left[\left\langle\varphi_l^n\left|\frac{d}{dx}\right|\varphi_{l'}^{(n-1)}\right\rangle a_{n-1,l'} + \left\langle\varphi_l^n\left|\frac{d}{dx}\right|\varphi_{l'}^{(n+1)}\right\rangle a_{n+1,l'}\right], \qquad (6)$$

$$a_{nl}(0) = \delta_{n0}\delta_{ll_0}.$$

On the left-hand side of Eq. (6) all the terms are expanded in $|n\omega/\varepsilon_\square \square 1|$. The relationship between the longitudinal and transverse motion of electrons mentioned in the Introduction governs the dependence of $\Omega^{(n)}$ on *n* and nonorthogonality of the functions $\varphi_l^{(n)}$ and $\varphi_{l'}^{(n\pm1)}$. The nonorthognality of these functions means that not only the transitions between the neighboring levels are allowed for, but also those between distant levels / and /'. The difference between the functions is governed by a small parameter $\omega/\varepsilon_\perp \ll 1$, but the matrix elements in the system of equations (6) have to be calculated rigorously bearing in mind that if $l \gg 1$ and $l\omega/\varepsilon_\perp \ll 1$, then $l \gg k, k'$, where $k = l - l_0$, and $k' = l' - l_0$.

The matrix elements in the system of equations (6) can be calculated by a method similar to that employed in Ref. [8], which finally gives

$$i\dot{a}_{nk} + \left[\frac{n\omega}{2\gamma^2} - k\Omega - \frac{(n\omega)^2}{2\varepsilon\gamma^2} + \frac{n\omega l_0 \Omega}{2\varepsilon} + \frac{nk\omega l_0 \Omega}{2\varepsilon} - \frac{3}{8}\frac{(n\omega)^2 l_0 \Omega}{\varepsilon^2}\right] a_{nk}$$
$$= i\mathrm{E}_{int} \sum_{k'} \left\{\left[\frac{J_{k-k'+1}(-z) + J_{k-k'-1}(-z)}{2}\right] a_{n-1,k'} + \left[\frac{J_{k'-k-1}(-z) + J_{k'-k'+1}(-z)}{2}\right] a_{n+1,k'}\right\}, \quad (7)$$

where $z = l_0\omega/4\varepsilon$, $\mathrm{E}_{int} = (eE_0/2\omega)(l_0\Omega/2\varepsilon)^{1/2}$ and the sum over $k'$ should include only the terms with the values of $k'$ which are of parity opposite to $k$.

In the resonance approximation, when

$$\omega \approx \omega_{res}^{(s)} = 2(2s+1)\gamma^2\Omega/(1+l_0\Omega\gamma^2/\varepsilon) \equiv (2s+1)\omega_{res}, \quad (8)$$

where $s = 0, 1, 2, \ldots$ we can substitute $k = (2s + 1)n$, and $k' = (2s + 1)(n \pm 1)$ in the system (2.4.7). Consequently, the equations for $a_n(t) \equiv a_{n,(2s+1)n}$ become identical with the equations known from the theory of a free-electron laser with an undulator [1]:

$$i\dot{a}_n - \mathrm{E}_{anh}(2n\Delta/\omega + n^2)a_n = \tilde{\mathrm{E}}_{int}(a_{n+1} + a_{n-1}), \quad a_n(0) = \delta_{n0}, \quad (9)$$

where

$$\tilde{\mathrm{E}}_{int} = (-1)^s \mathrm{E}_{int} F_s(z); \quad F_s(z) = J_s(z) - J_{s+1}(z), \quad (10)$$

and the following usual notation is introduced for the energy detuning $\Delta_s$ and the anharmonicity energy $\mathrm{E}_{anh}$:

$$\Delta_s = (\omega/2\mathrm{E}_{anh})\left[(2s+1)\Omega - 1/2\omega\left(1/\gamma^2 + l_0\Omega/\varepsilon\right)\right], \quad (11a)$$

$$\mathrm{E}_{anh} = (\omega/2\varepsilon)\left[\omega/\gamma^2 - (2s+1)\Omega + 3l_0\Omega\omega/4\varepsilon\right]. \quad (11b)$$

We shall finally obtain estimates of the approximations made on transition from the Klein-Gordon equation to the system (7) bearing in mind that $\langle x^2 \rangle \sim l_0/(\Omega\varepsilon)$, $\langle x^4 \rangle \sim l_0^2/(\Omega\varepsilon)^2$, $\partial/\partial t \sim 1/t$, and using in estimates the approximation $\varepsilon_\perp \sim m$. We can easily show that the relative smallness of the ignored terms proportional to $d^2 a_{nl}/dt^2$, $e\Phi(da_{nl}/dt)$ and $e^2 A_{em}^2 a_{nl}$ is governed by the small parameters $1/\varepsilon t$, $\varepsilon_\perp/\varepsilon \leq 1/\gamma$, and $eA_{em}/m\gamma^{1/2}$, respectively. The term proportional to $\Phi^2$ introduces into the system (9) a contributio $\sim (l^2\Omega^2/\varepsilon)a_n = \left[l_0^2\Omega^2/\varepsilon + 2l_0 n\Omega^2/\varepsilon + (n\Omega)^2/\varepsilon\right]a_n$. The n-independent part of this contribution is readily removed by a phase transformation. The part linear in *n* determines the correction to the detuning $\Delta_s$, which is small compared with $l_0 \omega n \Omega / \varepsilon$ in respect of the parameter $\Omega/\omega \ll 1$; the part quadratic in respect of *n* governs the correction to $E_{anh}$, which is small compared with $l_0 \Omega \omega^2 / \varepsilon^2$ in respect of the parameter $\varepsilon_\perp / \varepsilon \leq 1/\gamma \ll 1$.

## 3. CONCLUSIONS

Quantum-mechanical equations are derived for the amplitudes of the probabilities of transitions in a free-electron laser when transverse electron oscillations are due to a static potential, which is homogeneous in the direction of motion of the beam, but has a transverse gradient. The resonance frequency of the system is found

The main resonance frequency is shown to depend on the initial conditions of the electron, and in particular on its initial transversal coordinate $x_0$.

It seems also to be worthwhile to summarize the conditions under which the results described are applicabie. These conditions are given by inequalities by the assumptions $\alpha \ll 1$, and $x_0 \Omega \leq d_e / \lambda_0 \ll 1$, $\Omega T \gg 1$, and also by the condition $d_e \approx \alpha / \Omega$. The strongest restrictions result from the conditions $\alpha \ll 1$ and $\alpha \gamma \gg 1$ which give, for $d_e \approx \alpha / \Omega$

$$(\lambda_0 / 2\pi) > d_e > \lambda_0 / 2\pi \gamma.$$

We have ignored above both angle ($\Delta \alpha$ and energetic ($\Delta \varepsilon$) widths of the beam. This approximation is justified if

$$\Delta \alpha / \alpha, \quad \Delta \varepsilon / \varepsilon \ll \Delta \omega / \omega \ll 1/\Omega T.$$

The corresponding restrictions are not too strong because $\Omega T$ is really not too large, usually $\Omega T \approx 220$.